\documentstyle[12pt,epsf]{article}

\expandafter\ifx\csname amssym12.def\endcsname\relax \else\endinput\fi
\expandafter\edef\csname amssym12.def\endcsname{%
       \catcode`\noexpand\@=\the\catcode`\@\space}
\catcode`\@=11
\def\undefine#1{\let#1\undefined}
\def\newsymbol#1#2#3#4#5{\let\next@\relax
 \ifnum#2=\@ne\let\next@\msafam@\else
 \ifnum#2=\tw@\let\next@\msbfam@\fi\fi
 \mathchardef#1="#3\next@#4#5}
\def\mathhexbox@#1#2#3{\relax
 \ifmmode\mathpalette{}{\m@th\mathchar"#1#2#3}%
 \else\leavevmode\hbox{$\m@th\mathchar"#1#2#3$}\fi}
\def\hexnumber@#1{\ifcase#1 0\or 1\or 2\or 3\or 4\or 5\or 6\or 7\or 8\or
 9\or A\or B\or C\or D\or E\or F\fi}

\font\tenmsa=msam10 scaled\magstep1
\font\sevenmsa=msam7 scaled\magstep1
\font\fivemsa=msam5 scaled\magstep1
\newfam\msafam
\textfont\msafam=\tenmsa
\scriptfont\msafam=\sevenmsa
\scriptscriptfont\msafam=\fivemsa
\edef\msafam@{\hexnumber@\msafam}
\mathchardef\dabar@"0\msafam@39
\def\dashrightarrow{\mathrel{\dabar@\dabar@\mathchar"0\msafam@4B}}
\def\dashleftarrow{\mathrel{\mathchar"0\msafam@4C\dabar@\dabar@}}

\def\ulcorner{\delimiter"4\msafam@70\msafam@70 }
\def\urcorner{\delimiter"5\msafam@71\msafam@71 }
\def\llcorner{\delimiter"4\msafam@78\msafam@78 }
\def\lrcorner{\delimiter"5\msafam@79\msafam@79 }
\def\yen{{\mathhexbox@\msafam@55 }}
\def\checkmark{{\mathhexbox@\msafam@58 }}
\def\circledR{{\mathhexbox@\msafam@72 }}
\def\maltese{{\mathhexbox@\msafam@7A }}

\font\tenmsb=msbm10 scaled\magstep1
\font\sevenmsb=msbm7 scaled\magstep1
\font\fivemsb=msbm5 scaled\magstep1
\newfam\msbfam
\textfont\msbfam=\tenmsb
\scriptfont\msbfam=\sevenmsb
\scriptscriptfont\msbfam=\fivemsb
\edef\msbfam@{\hexnumber@\msbfam}
\def\Bbb#1{{\fam\msbfam\relax#1}}
\def\widehat#1{\setbox\z@\hbox{$\m@th#1$}%
 \ifdim\wd\z@>\tw@ em\mathaccent"0\msbfam@5B{#1}%
 \else\mathaccent"0362{#1}\fi}
\def\widetilde#1{\setbox\z@\hbox{$\m@th#1$}%
 \ifdim\wd\z@>\tw@ em\mathaccent"0\msbfam@5D{#1}%
 \else\mathaccent"0365{#1}\fi}
\font\teneufm=eufm10 scaled\magstep1
\font\seveneufm=eufm7 scaled\magstep1
\font\fiveeufm=eufm5 scaled\magstep1
\newfam\eufmfam
\textfont\eufmfam=\teneufm
\scriptfont\eufmfam=\seveneufm
\scriptscriptfont\eufmfam=\fiveeufm

\csname amssym.def\endcsname

\newif{\ifcomentarios}
\comentariosfalse
\renewcommand{\theequation}{\thesection.\arabic{equation}}
\newtheorem{theorem}{Theorem}
\newtheorem{lemma}[theorem]{Lemma}

\newtheorem{proposition}[theorem]{Proposition}

\newtheorem{remark}[theorem]{Remark}

\newcommand{\zerarcounters}
{
\setcounter{equation}{0}
\setcounter{theorem}{0}
}

\newcommand{\Fullbox}{\bigskip\hfill{\rule{2.5mm}{2.5mm}}}

\newcommand{{\fa}}{(\phi_1\times\cdots\times\phi_n )_{in}}
\newcommand{{\fb}}{(\phi_2\times\cdots\times\phi_n )_{in}}
\newcommand{{\fc}}{(\phi_{m+1}\times\cdots\times\phi_n )_{in}}
\newcommand{{\fd}}{(\phi_{n-k+1}\times\cdots\times\phi_n )_{in}}
\newcommand{{\fe}}{(\phi_{n-k}\times\cdots\times\phi_n )_{in}}
\newcommand{{\ff}}{(\phi_{n-k+1}\times\cdots\times\phi_n )_{in}}
\newcommand{{\fg}}{(\phi_{n-k}\times\cdots\times\phi_n )_{in}}
\newcommand{{\fh}}{(\phi_{1}\times\cdots\times\phi_k )_{out}}
\newcommand{{\fz}}{(\phi_{1}\times\cdots\times\phi_m )_{out}}
\newcommand{{\fl}}{(\phi_2\times\cdots\times\phi_n )_{out}}
\newcommand{{\fm}}{\phi_{1,\ldots , n}}
\newcommand{{\fn}}{\phi_{2,\ldots , n}}

\newcommand{\calB}{{\cal B}}

\newcommand{\calF}{{\cal F}}

\newcommand{\calH}{{\cal H}}

\newcommand{\calO}{{\cal O}}

\newcommand{\calZ}{{\cal Z}}

\newcommand{\be}{\begin{equation}}
\newcommand{\ee}{\end{equation}}
\newcommand{\bma}{\begin{displaymath}}
\newcommand{\ema}{\end{displaymath}}
\newcommand{\bc}{\begin{center}}
\newcommand{\ec}{\end{center}}

\newcommand{\om}{{\omega }}

\newcommand{\Om}{\Omega}

\newcommand{\bear}{\begin{eqnarray}}
\newcommand{\eear}{\end{eqnarray}}

\newcommand{\uflex}
{{\scriptstyle {\raise 9pt\hbox{$\backslash$}\,\!\!\!\!\!\Bigg\vert}}}

\newcommand{\N}{\Bbb N}
\newcommand{\R}{\Bbb R}
\newcommand{\C}{\Bbb C}
\newcommand{\E}{\Bbb E}

\newcommand{\ncm}{\newcommand}
\ncm{\rncm}{\renewcommand}
\ncm{\id}{{\bf 1}}
\ncm{\beq}{\begin{equation}}
\ncm{\eeq}{\end{equation}}
\ncm{\bea}{\begin{eqnarray}}
\ncm{\beanon}{\begin{eqnarray*}}
\ncm{\eea}{\end{eqnarray}}
\ncm{\eeanon}{\end{eqnarray*}}
\ncm{\fns}{\footnotesize}
\ncm{\setc}[1]{\setcounter{equation}{#1}}
\newcounter{eqnr}
\newenvironment{eqnarrayabc}{\stepcounter{equation}
  \setcounter{eqnr}{\value{equation}}\setc{0}
  \rncm{\theequation}{\thesection.\arabic{eqnr}\alph{equation}}
  \begin{eqnarray}}{\end{eqnarray}\setc{\value{eqnr}}}
\ncm{\eqboxabc}[3]{\newline\parbox[t]{1.5cm}{#1}\hfill
  \parbox[b]{12cm}{\begin{eqnarray*} #3\end{eqnarray*}}\hfill
   \parbox[b]{1.5cm}{\vspace{-0.0cm}\begin{eqnarrayabc}#2\end{eqnarrayabc}}\newline}
\ncm{\eqbox}[2]{\newline\parbox{1.5cm}{#1}\hfill
  \parbox{12cm}{\beanon #2\eeanon}\hfill
  \parbox{1cm}{\bea\eea}\newline}
\ncm{\nr}[1]{\parbox{1cm}{\begin{eqnarrayabc}#1\end{eqnarrayabc}}\\}

\ncm{\kal}[1]{\mbox{$\cal #1 $}}
\ncm{\mrk}[1]{\!\!\! #1 \!\!\!} 
\ncm{\qed}{\hspace*{0.4cm}\rule{0.24cm}{0.24cm}}  
\ncm{\mbold}[1]{\mbox{\boldmath $ #1 $}}   
\ncm{\bm}{\mbold}
\ncm{\str}{\stackrel}
\ncm{\sub}{\subset}
\ncm{\e}{\varepsilon}
\ncm{\ka}{\kappa}
\ncm{\inputc}[1]{\begin{center}\input{#1}\end{center}}
\ncm{\lto}{\longrightarrow}
\ncm{\x}{\times}
\ncm{\bmm}{\bm{\cal M}}
\ncm{\cp}{{\bf P}}    
\ncm{\bfp}{{\bf P}}
\ncm{\bmi}{\bm{i}}
\ncm{\bmom}{\bm{\om}}
\ncm{\bmOm}{\bm{\Om}}
\ncm{\res}{\restriction}
\ncm{\bmL}{\bm{\cal L}}
\ncm{\bmell}{\bm{\ell}}
\ncm{\bmE}{\bm{\cal E}}
\ncm{\bme}{\bm{e}}
\ncm{\bmpi}{\bm{\pi}}
\ncm{\bmr}{\bm{r}}
\ncm{\bmsigma}{\bm{\sigma}}
\ncm{\wt}{\widetilde}

\newcommand{\text}{\rm}

\newcommand{\dfrac}{\displaystyle\frac}
\newcommand{\dsum}{\displaystyle\sum}
\newcommand{\dprod}{\displaystyle\prod}

\newcommand{\EndofStatement}{\samepage\bigskip\hfill\Box}

\newcommand{\beaa}{\begin{eqnarray}}
\newcommand{\eeaa}{\end{eqnarray}} 

\newcommand{\Proof}{\noindent Proof}

\begin{document}

\author{{\bf J. C. A. Barata}\thanks{
e-mail: jbarata{@}fma.if.usp.br. Partially supported by CNPq.} \, \& \, {\bf 
D. H. U. Marchetti}\thanks{
e-mail: marchett{@}ime.usp.br. Partially supported by CNPq.} \\
Instituto de F\'\i sica \\
Universidade de S\~ao Paulo\\
P. O. Box 66 318\\
S\~ao Paulo, SP, Brazil}
\title{
The Two-Point Function and the Effective Magnetic Field in 
Diluted Ising Models on the Cayley Tree
}
\date{}
\maketitle

\begin{abstract} 
  Some results on the two-point function and on the analytic structure
  of the momenta of the effective fugacity at the origin for a class
  of diluted ferromagnetic Ising models on the Cayley tree are
  presented.
\end{abstract}


\tableofcontents

\newpage

\section{Introduction and Previous Results}
\label{Introduction}

\zerarcounters

In a previous work \cite{BM1} a detailed analysis of the presence of
Griffiths' singularities has been performed in a class of diluted
ferromagnetic Ising models on the Cayley tree. This involved the study
of the analytic structure of the quenched magnetization at the origin,
as a function of the fugacity, as well as the study of its
differentiability for real values of the magnetic field. A detailed proof
has been presented of the infinite differentiability of the quenched
magnetization for real magnetic fields under suitable choices of
disorder. The set of Lee-Yang zeros of the deterministic ferromagnetic
Ising model on the Cayley tree has been also studied and detailed
results have been presented, both in the ferromagnetic and in the
paramagnetic phases, showing that this set becomes dense in subsets of
the unit circle when the thermodynamic limit is taken. In the
ferromagnetic phase, in particular, the set of Lee-Yang zeros becomes
dense in the whole unit circle when this limit is performed.  These
facts are responsible for the emergence of Griffiths' singularities
and for the absence of metastable states in the diluted models
considered.

In this note we extend our previous results on the diluted models
considered. Our main intention here is to present some results on the
momenta of the effective fugacity (defined below) produced on the spin
at the origin in the random model introduced in \cite{BM1}, as well as
an explicit expression for the two-point functions of that model. Some
conclusions on the quenched susceptibility at the origin are drawn.

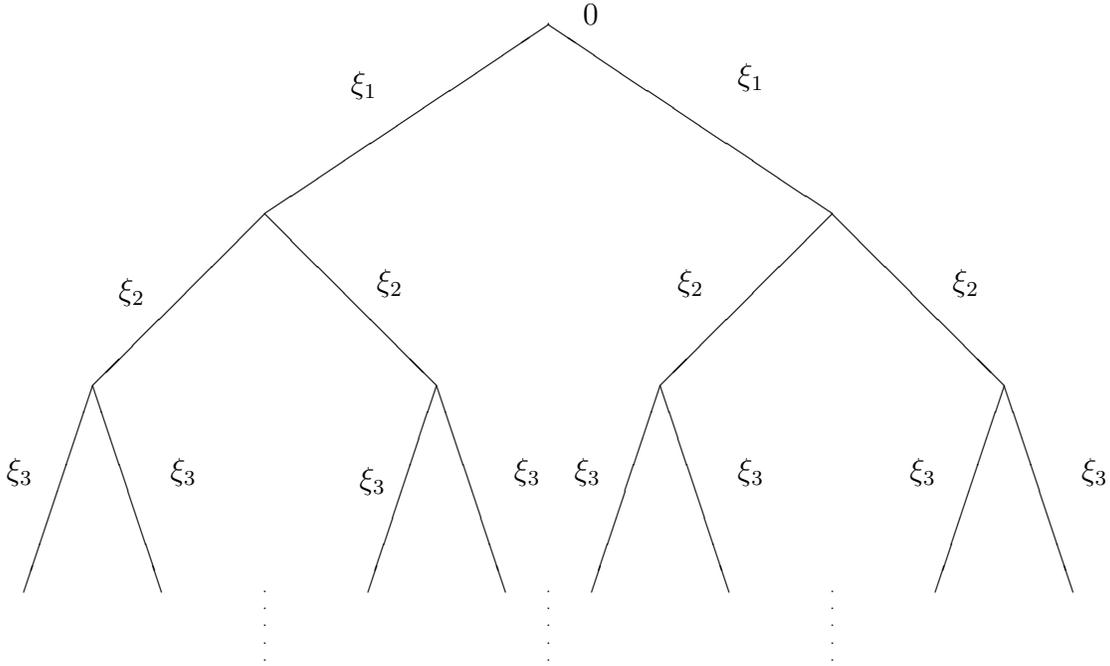
\begin{figure}[hbtp]
\begin{center}

\setlength{\unitlength}{0.009in}%
\begingroup\makeatletter\ifx\SetFigFont\undefined
\def\x#1#2#3#4#5#6#7\relax{\def\x{#1#2#3#4#5#6}}%
\expandafter\x\fmtname xxxxxx\relax \def\y{splain}%
\ifx\x\y   
\gdef\SetFigFont#1#2#3{%
  \ifnum #1<17\tiny\else \ifnum #1<20\small\else
  \ifnum #1<24\normalsize\else \ifnum #1<29\large\else
  \ifnum #1<34\Large\else \ifnum #1<41\LARGE\else
     \huge\fi\fi\fi\fi\fi\fi
  \csname #3\endcsname}%
\else
\gdef\SetFigFont#1#2#3{\begingroup
  \count@#1\relax \ifnum 25<\count@\count@25\fi
  \def\x{\endgroup\@setsize\SetFigFont{#2pt}}%
  \expandafter\x
    \csname \romannumeral\the\count@ pt\expandafter\endcsname
    \csname @\romannumeral\the\count@ pt\endcsname
  \csname #3\endcsname}%
\fi
\fi\endgroup
\begin{picture}(625,385)(90,380)
\thinlines
\put(240,640){\line( 3, 2){165}}
\put(405,750){\makebox(0.1111,0.7778){\SetFigFont{5}{6}{rm}.}}
\put(570,640){\line(-3, 2){165}}
\put(240,640){\line(-1,-1){100}}
\put(240,640){\line( 1,-1){100}}
\put(570,640){\line( 1,-1){100}}
\put(570,640){\line(-1,-1){100}}
\put(140,540){\line(-1,-3){ 40}}
\put(140,540){\line( 1,-3){ 40}}
\put(340,540){\line( 1,-3){ 40}}
\put(470,540){\line( 1,-3){ 40}}
\put(670,540){\line( 1,-3){ 40}}
\put(340,540){\line(-1,-3){ 40}}
\put(470,540){\line(-1,-3){ 40}}
\put(670,540){\line(-1,-3){ 40}}
\multiput(240,420)(0.00000,-10.00000){5}{\makebox(0.1111,0.7778){\SetFigFont{5}{6}{rm}.}}
\multiput(405,420)(0.00000,-10.00000){5}{\makebox(0.1111,0.7778){\SetFigFont{5}{6}{rm}.}}
\multiput(570,420)(0.00000,-10.00000){5}{\makebox(0.1111,0.7778){\SetFigFont{5}{6}{rm}.}}
\put(425,750){\makebox(0,0)[lb]{\smash{\SetFigFont{12}{14.4}{rm}$0$}}}
\put(290,710){\makebox(0,0)[lb]{\smash{\SetFigFont{12}{14.4}{rm}$\xi_1$}}}
\put(515,715){\makebox(0,0)[lb]{\smash{\SetFigFont{12}{14.4}{rm}$\xi_1$}}}
\put(155,590){\makebox(0,0)[lb]{\smash{\SetFigFont{12}{14.4}{rm}$\xi_2$}}}
\put(305,595){\makebox(0,0)[lb]{\smash{\SetFigFont{12}{14.4}{rm}$\xi_2$}}}
\put(640,595){\makebox(0,0)[lb]{\smash{\SetFigFont{12}{14.4}{rm}$\xi_2$}}}
\put(295,480){\makebox(0,0)[lb]{\smash{\SetFigFont{12}{14.4}{rm}$\xi_3$}}}
\put(385,485){\makebox(0,0)[lb]{\smash{\SetFigFont{12}{14.4}{rm}$\xi_3$}}}
\put(615,485){\makebox(0,0)[lb]{\smash{\SetFigFont{12}{14.4}{rm}$\xi_3$}}}
\put(715,485){\makebox(0,0)[lb]{\smash{\SetFigFont{12}{14.4}{rm}$\xi_3$}}}
\put(515,485){\makebox(0,0)[lb]{\smash{\SetFigFont{12}{14.4}{rm}$\xi_3$}}}
\put(185,485){\makebox(0,0)[lb]{\smash{\SetFigFont{12}{14.4}{rm}$\xi_3$}}}
\put( 90,485){\makebox(0,0)[lb]{\smash{\SetFigFont{12}{14.4}{rm}$\xi_3$}}}
\put(480,595){\makebox(0,0)[lb]{\smash{\SetFigFont{12}{14.4}{rm}$\xi_2$}}}
\put(420,485){\makebox(0,0)[lb]{\smash{\SetFigFont{12}{14.4}{rm}$\xi_3$}}}
\end{picture}

\end{center}
\caption{The homogeneous rooted Cayley tree of order $2$ and the
  couplings $\xi_M$.}
\label{ArvoredeCayley}
\end{figure}

As in \cite{BM1}, we consider a random Ising ferromagnet in an homogeneous
rooted Cayley tree  of order $2$, ${\cal C}_2$,
described by the Hamiltonian
\begin{equation}
\label{H}
\calH (\sigma ;\;\xi ):=-\sum_{\langle xy\rangle }J_{xy}\ \sigma
_x\sigma _y- H \sum_x\sigma _x 
\end{equation}
where $\sigma :{\cal C}_2 \ni x\longmapsto \sigma _x\in \left\{ 1,-1\right\} $ is
a configuration of spins. 
The coupling constants $J_{xy}$ are random variables 
whose distribution is described as follows.
To each generation $M$ of ${\cal C}_2$ a Bernoulli random variable $\xi _M$ 
\begin{equation}
\label{xim}\xi _M=\left\{ 
\begin{array}{lll}
1, &  & \mbox{ with probability } p_M 
\\ 
0, &  & \mbox{ with probability } q_M=1-p_M 
\end{array}
\right. 
\end{equation}
is assigned and we then set $J_{xy}=$ $\xi _M$ if $\langle xy\rangle \equiv
b $ is a bond at the generation $M$ and $0$ otherwise. 
See Figure \ref{ArvoredeCayley}.

For any event $A$, which is an element of the $\sigma $-algebra generated by
the space of spin configurations, we define the quenched expected
value of $A $ as 
\begin{equation}
\label{Abar}\overline{A}\equiv {\E}[A]={\E}_\xi \left\langle A\right\rangle
(\xi ) 
\end{equation}
where ${\E}_\xi $ is the expectation with respect to the disorder variables
and 
\begin{equation}
\label{Axi}\left\langle A\right\rangle (\xi )=\frac 1{Z(\xi )}\sum_\sigma
A(\sigma )\ e^{-\beta \calH(\sigma ;\;\xi )} 
\end{equation}
is the thermal average with 
$$
Z(\xi )=\sum_\sigma \ e^{-\beta \calH(\sigma ;\;\xi )} 
$$
being the partition function.

The expectation (\ref{Abar}) is a function of the inverse temperature $\beta 
$, the external magnetic field $H$ and the sequence of Bernoulli
parameters 
${\bf p}=\left\{ p_j\right\} _{j=1}^\infty $. We let $\pi =\left\{ {\bf p} :\;
0<p_j\leq 1,\ j=1,\, 2,\ldots \right\} $ and for $0\leq a\leq 1$, let $\pi _a=$ $\left\{ {\bf p\in \pi }:\; \lim
\limits_{n\rightarrow \infty }\ p_1\ldots p_n=a\right\} $. Also, for
convenience, we write $\zeta =e^{-2\beta }$ and $z=e^{-2\beta H}$, the
so-called fugacity.

We first take the expectation in a finite tree ${\cal C}_{2,\, M}$,
$M=1,2,\ldots $, with free boundary conditions and study the function 
$\overline{A}: \; [0, \, 1]\times {\C} \times \pi $ $\longmapsto \overline{A}
(\zeta ,\, z,\, {\bf p})\in {\C}$ in the limit as $M$ goes to
infinite. 

In \cite{BM1} 
we were concerned, following \cite{G},
with the magnetization at the 
origin $m:={\E}[\sigma _0]={\E}_\xi \langle \sigma
_0\rangle (\xi )$ which can be written as 
\begin{equation}
\label{m}m\;\;=\;\;F\;+\;I, 
\end{equation}
where 
$$
F\,:=\,\dsum_{M=0}^{\infty}a_M\ \langle \sigma _0\rangle _M(1)
$$
and
$$
I:= \lim\limits_{N\rightarrow \infty }p_1\ldots p_N\;\langle \sigma _0\rangle
_N(1)
$$ 
give the contribution to $m$ due to finite clusters and
the (unique) infinite cluster, respectively. Here, for any
observable $A$, $\langle A\rangle _M$ is the thermal average (\ref{Axi})
with ${\cal C}_2$ replaced by the finite tree ${\cal C}_{2,\, M}$ and
$a_M$, $M\in \N$, is given by 
$$
a_N  :=  p_1\ldots p_N \; (1-p_{N+1}),\qquad N\geq 1
$$
with $a_0  :=  1-p_1$. Clearly, $\{a_n\}_{n=0}^\infty $ is a summable
sequence with $\dsum_{n=0}^\infty a_n=1-a $
for $p\in \pi _a$.

Equation (\ref{m}) is a consequence of the fact that $\langle \sigma
_0\rangle (\xi )=\langle \sigma _0\rangle _M(1)$ for all $\xi $ such
that $\xi _1=\cdots =\xi _M=1$ and $\xi _{M+1}=0$ with the finite volume
one--point function given as follows (see Section
\ref{ThermalExpectationFormulas} for details):

\begin{equation}
\label{sigma2}
\left\langle \sigma _0\right\rangle _M(1)=\frac{1-z\ \tau
_z^{(M)}(1)}{1+z\ \tau _z^{(M)}(1)}
\end{equation}
where $\tau^{(n)}_z$ denotes the $n$-th composition of $\tau_z$ with
itself: 
$
\displaystyle
\tau^{(n)}_z := \underbrace{\tau_z \circ \cdots \circ \tau_z}_{n - {\rm times}}
$ 
and, for $z\in \C$, $\tau _z:\C \mapsto \C$ is a rational map given by
$
\tau _z(u):=h(zu), 
$
with  
\begin{equation}
\label{h}h(w):=\left( \frac{\zeta +w}{1+\zeta w}\right) ^2.
\end{equation}

The analytic properties of $F$ and $I$ have been studied in
\cite{BM1} based on the following facts.

Let us denote by $D_{<1}$ the open unit disk in the Riemann sphere
$S^2$ and by $D_{>1}$ the complement of its closure. In \cite{BM1} it
was shown that, for the whole interval $0\leq \zeta <1$, the function
$h$ maps $D_{<1}$, $D_{>1}$ and $S^1 $ into itself. This implies that,
as function of $z$, $\tau _z^{(n)}(1)$ maps each of the sets $D_{<1}$,
$S^1 $ and $D_{>1}$ into itself for all $n\in \N$ and all $\zeta \in
[0,\;1)$.  In addition, each of the sets $[0,\,1)$ and $(1,\,\infty
)\subset \R_{+}$ is also mapped into itself and, for $z=1$ one has
$\tau _1^{(n)}(1)=1$, for all $n\in \N$. The description of the Lee--Yang
singularities and the smoothness of the quenched magnetization are also
obtained from these and further properties of the orbit $\tau _z^{(n)}(1),
\; n\in \N$, of the dynamical system induced by $\tau _z$.
For the present analysis these properties will play an important
role as well.

Let $H_j$ be the effective magnetic 
field acting on the spins at generation $j$ and let $\Delta_j= e^{-2\beta
H_j}$ be the associated effective fugacity. We have the following recurrence
relation 
\begin{equation}
\label{Deltaj}\Delta _{j-1}=\tau _{j,\,z}(\Delta _j),\qquad j=1,\ldots ,M,
\end{equation}
with $\Delta _M=1$ and $z\in {\C}$, where 
\begin{equation}
\label{tauj}\tau _{j,\,z}(u)=h_j(zu)
\end{equation}
and $h_j$ is given by (\ref{h}) with $\zeta $ replaced by $\zeta
_j=e^{-2\beta \xi _j}$. In this work we will be concerned with
the effective fugacity at the origin $\Delta_0 = \tau^{(M)}_{j, \,z}
(1)$. The reason for this nomenclature comes from the analogy between
(\ref{sigma2}) 
and the expression for the magnetization of a lattice containing a
single spin: $ \displaystyle m = \frac{1 -z}{1 + z} $.

Since the couplings are random, so are the quantities $\Delta _j$'s. 
Note that $\Delta_j$ depends on the generation level $M$ of the Cayley
tree through its initial condition.  
The distribution of $\Delta_0 $, when $M\rightarrow \infty $, is worth
studying by its own right. In the next theorem 
the analytic properties of the moments ${\cal M}_s={\E}_\xi \left[
  \Delta _0(\xi )^s\right] $, $s\in {\N}$, of $\Delta _0$ 
will be described. We begin by noting that they can be written as

\begin{equation}
\label{Ms}{\cal M}_s\;=\;{\cal F}_s\;+\;{\cal I}_s, 
\end{equation}
where ${\cal F}_s\;:=\;\lim \limits_{M\rightarrow \infty }{\cal F}_{s,\;M}$
with 
\begin{equation}
\label{calF}{\cal F}_{s,M}\;:=\;\sum_{n=0}^{M-1}a_n\ \left( \tau
_z^{(n)}(1)\right) ^s 
\end{equation}
and ${\cal I}_s\;:=\;\lim \limits_{M\rightarrow \infty }{\cal I}_{s,\;M}$
with 
\begin{equation}
\label{calI}{\cal I}_{s,M}\;:=\;p_1\ldots p_M\;\left( \tau
_z^{(M)}(1)\right) ^s. 
\end{equation}

Our first result is in the following theorem, which  will be proven in
Section \ref{TheAnalyticityDomainofcalM}. The proof involves a careful 
analysis of the dynamical system induced by $ \tau_z$ on the complex
plane. 

\begin{theorem}[Moments] 
  {\label{moments}} For fixed $s\in {\N}$, let ${\cal M}_s:\;\C \times
  [0,\, 1]\times \pi \ni (z, \, \zeta ,\, {\bf p})\longmapsto {\cal M}_s(z,\, \zeta
  ,\, {\bf p})\in {\C}$ be given by (\ref{Ms}).  Then ${\cal M}_s$ is an
  analytic function in $z$ for $\left| z\right| <1$. In addition, for
  $\zeta \in [0,\;1]$, $ {\cal M}_s$ is a holomorphic function of $z$
  in the open set $\left\{ z\in \C ;\;\;\left| z\right| >1\right\}
  \setminus \overline{{\cal Z}}$, where $\overline{{\cal Z}}$ is the
  closure of
  $$
  {\cal Z}:=\bigcup_{a\in \N}{\cal Z}_a ,
  $$
  with ${\cal Z}_a$ being the set of the $2^{a+1}-1$ solutions of
  the equation
  $$
  \underbrace{\tau _z\circ \cdots \circ
    \tau_z}_{a\text{-times}}(1)+ \frac 1{\zeta z}=0.
  $$
  Furthermore, ${\calZ}_a$ are self-conjugated sets (i.e.,
  $\overline{z}\in {\cal Z}_a$ if $z\in {\cal Z}_a$) satisfying ${\cal
    Z}_a\cap \left[ 0,\infty \right)=\emptyset $ and ${\cal Z}^0\equiv
  \overline{{\cal Z}}\setminus {\cal Z}\subset S^1$.  There exists a
  $\zeta_0\in \left[ 0, \, 1/3\right]$ with $\zeta_0\simeq
  0.29559$\footnote{See also Remark \ref{zetazero}.} such that, for
  $\zeta \in [0,\;\zeta _0)$, we have the inclusion
  \be
  {\calZ}_a\subset \left\{ z\in {\C}:1<\left| z\right| < \left(
      1/\zeta \right)^{1/(a+1)}\right\} .
  \label{acumulacaodosZs}
  \ee
  We conjecture that ${\cal  Z}^0=S^1$ for $\zeta \in \left[ 0,\, 1/3\right] $.
  $\EndofStatement$
\end{theorem}

\begin{figure}[hbtp]
\epsfysize=8cm
\begin{center}
\leavevmode\epsfbox{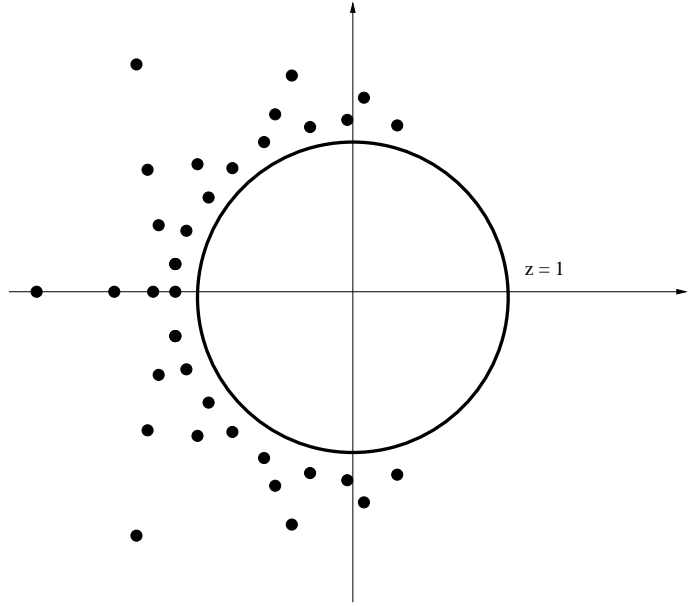}
\end{center}
\caption{The location of the set of poles $\calZ$ and the unit circle 
  $S^1$. The points of $\calZ$ accumulate on subsets ${\cal Z}^0$ of the
  unit circle. Numerical computations indicate that ${\cal Z}^0=S^1$
  in the ferromagnetic phase. }
\label{graficodosPolos}
\end{figure}

The elements of the set ${\cal Z}_a $ are poles of
$\tau^{(a+1)}_{z}(1)$.  These singularities are removable in the
magnetization (\ref{sigma2}).  Theorem \ref{moments} states that the
set of singular points of $ {\cal M}_s$ is located in $D_{>1}$ and
accumulate on a subset ${\cal Z}^0$ of $S^1$. Figure
\ref{graficodosPolos} describes the singular set $\calZ$. We
conjecture that the accumulation set ${\cal Z}^0$ coincides with the
accumulation set of the Lee-Yang singularities of the magnetization at
the origin, studied in \cite{BM1}. Numerical computations seem to
confirm this idea. The inclusion (\ref{acumulacaodosZs}) indicates how
fast the sets ${\calZ}_a $ tend to accumulate on $S^1$ for $\zeta \in
[0,\;\zeta _0)$.

As a consequence of the recurrence relation (\ref{Deltaj}), we are also able
to compute the two-point function explicitly. The next result shows that  the
fluctuation on the Cayley tree is bounded by that of the one-dimensional
lattice model.

\begin{theorem}[Two-Point Function]
{\label{tpf} } The truncated two-point function 
$$
\left\langle \sigma _0;\ \sigma _x\right\rangle _N(\xi ):=\left\langle
\sigma _0\ \sigma _x\right\rangle _N(\xi )-\left\langle \sigma
_0\right\rangle _N(\xi )\ \left\langle \sigma _x\right\rangle _N(\xi ) 
$$
can be written as
\begin{equation}
\label{t2pf}\left\langle \sigma _0;\ \sigma _x\right\rangle _N(\xi
)=s(z\Delta _0)\ \prod_{j=1}^nt(\zeta _j,z\Delta _j)
\end{equation}
where $n=n(x)=$\mbox{dist}$(0,\,x)$ is the generation of the site $x$,
$$
s(x):=1-\left( \frac{1-x}{1+x}\right) ^2 
$$
and 
$$
t(x)=t(\zeta ,x):=\frac{\zeta ^{-1}-\zeta }{x^{-1}+x+\zeta ^{-1}+\zeta }. 
$$

For all $\zeta \in \left[ 0,\, 1\right] $, $z\in \R_{+}$ and ${\bf p}\in \pi _a$
, $0\leq a\leq 1$, the following identity holds for the expected value ${\E}
_\xi \left\langle \sigma _0;\ \sigma _x\right\rangle (\xi )=\lim
\limits_{N\rightarrow \infty }{\E}_\xi \left\langle \sigma _0;\ \sigma
_x\right\rangle _N(\xi )$: 
\begin{equation}
\label{expect}{\E}_\xi \left\langle \sigma _0;\ \sigma _x\right\rangle (\xi
)=\sum_{k=n}^\infty a_k\ s(w_k)\ \left[ \prod_{j=1}^nt(w_{k-j})\right] +a\
s(w)\ \left[ t(w)\right] ^n
\end{equation}
where $w=w(\zeta ,z)$ is the limit point of the sequence
$w_n=z\, \tau _z^{(n)}(1),
\; n\in {\N}$.
$\EndofStatement$
\end{theorem}

\begin{remark}
A simple upper bound on (\ref{expect}) for the paramagnetic phase can be
obtained by using Perez's self-avoiding random walk estimate \cite{P}. Only
one term, corresponding to the single self-avoiding path connecting $0$ to $x
$, contributes to the correlation functions in ${\cal C}_k$. This fact shows
that the Ising model on the Cayley tree behaves, on what concerns the
asymptotic behavior of the correlations, as a one dimensional system. As a
consequence, the correlation length is always finite, even at the transition
point. (Note that $\left| t(\zeta ,x)\right| \leq 1/2$ if $\zeta \in \left[
1/3,\, 1\right] $ and $x\in \R_{+}$).
\end{remark}

Our last results shows that the quenched susceptibility at the origin $\chi
:=\lim \limits_{N\rightarrow \infty }\chi _N$ given by 
\begin{equation}
\label{susc}\chi _N:=\sum_{x\in {\cal C}_{2,N}}{\E}_\xi \left\langle \sigma
_0;\ \sigma _x\right\rangle (\xi )
\end{equation}
diverges as $\left| \zeta -\zeta _c\right| ^\gamma $ at the critical
point 
$\zeta _c=1/3$ with an exponent given by the classical theory $\gamma =1$.
More precisely

\begin{theorem}[Susceptibility]
\label{suscept}The quenched susceptibility at origin $\chi=\chi (\zeta , \, z) $ is finite for
all $\zeta \in \left[ 0,\, 1\right] $, $z\in \R_{+}\left\backslash \{1\}\right. 
$ and ${\bf p}\in \pi $. Moreover, $\lim \limits_{z\rightarrow 1}\chi $ is
also finite provided $\zeta \neq \zeta _c$.

In addition, for ${\bf p}\in \pi _a$, $0<a\leq 1$, the following asymptotic
behavior 
\begin{equation}
\label{asympt} \chi(\zeta , \, 1) \sim C_\eta \left| \eta
\right| ^\gamma \qquad \text{\mbox{as} }\qquad \eta :=\zeta -\zeta
_c\rightarrow 0
\end{equation}
holds with $\gamma =1$, $\lim _{\eta \searrow 0}C_\eta =4a/3$ and $\lim
_{\eta \nearrow 0}C_\eta =2a/3$ provided the condition 
\begin{equation}
\label{summcond}\sum_{n=0}^\infty \,n\,a_n<\infty 
\end{equation}
is satisfied. 
$\EndofStatement$
\end{theorem}

The proofs of Theorems \ref{tpf} and \ref{suscept} will be given in Section 
\ref{ThermalExpectationFormulas}.




\section{The Two-Point Correlation Function}
\zerarcounters
\label{ThermalExpectationFormulas}

This section is dedicated to the proof of Theorems \ref{tpf} and
\ref{suscept}. 
We shall consider the truncated two-point function 
\begin{equation}
\label{t2pfM}
\left\langle \sigma _0;\sigma _x\right\rangle _M(\xi
)=\left\langle \sigma _0\sigma _x\right\rangle _M(\xi )-\left\langle
  \sigma_0
\right\rangle _M(\xi )\ \left\langle \sigma _x\right\rangle_M(\xi )
\end{equation}
where 
\begin{equation}
\label{cf}\left\langle \sigma _0\sigma _x\right\rangle _M(\xi )=\frac
1{Z_M(\xi )}\sum_\sigma \sigma _0\ \sigma _x\ e^{-\beta H(\sigma ;\;\xi )}
\end{equation}
for all $x\in {\cal C}_{2,\,M}$ with $M$ large enough.


\subsection{The One Point Function}

Let us first recall the iteration procedure leading to expression
(\ref{sigma2}) for the magnetization at the origin. We start by
computing the partition function $Z_M(\xi )$ in a finite tree with $M$
generations.

Let ${Z}_j=({Z}_j^{+},{Z}_j^{-})$ , $j=0,\;1,\ldots
,M$, be a sequence of two-component vectors defined recursively by

\begin{equation}
\label{Zhat}
\begin{array}{lll}
{Z}_{j-1}^\sigma  & := & \left( \dsum\limits_{\sigma ^{\prime }=\pm
1}e^{\beta \xi _j\sigma \sigma ^{\prime }}\ e^{\beta h\sigma ^{\prime }}\ 
{Z}_j^{\sigma ^{\prime }}\right) ^2 \\  &  &  \\  
& = & \left( \zeta _j z\right) ^{-1}\left( \zeta _j ^{(1-\sigma )/2}\ 
{Z}_j^{+}+\zeta _j ^{(1+\sigma )/2}\ z\ {Z}_j^{-}\right) ^2
\end{array}
\end{equation}
with ${Z}_M^{+}={Z}_M^{-}=1$.

If the spin configurations are summed starting from the branches towards the
root, the partition function $Z_M(\xi )$ 
can be written as 
\begin{equation}
\label{ZM}Z_M(\xi)=z^{-1/2}\ {Z}_0^{+}+z^{1/2}\ {Z}_0^{-} 
\end{equation}

To compute the one point-function 
\begin{equation}
\label{sigma0}\left\langle \sigma _0\right\rangle _M(\xi )=\frac 1{Z_M(\xi
)}\sum_\sigma \sigma _0\ e^{-\beta H(\sigma ;\;\xi )},
\end{equation}
we repeat the procedure leading to (\ref{ZM}) for the numerator in (\ref
{sigma0}). Except by the last summation on the spin at the root, all
remaining ones give exactly the previous expressions. We thus have 
\begin{equation}
\label{sigma1}\left\langle \sigma _0\right\rangle _M(\xi)=\frac{z^{-1/2}\ 
{Z}_0^{+}-z^{1/2}\ {Z}_0^{-}}{z^{-1/2}\ 
{Z}_0^{+}+z^{1/2}\ 
{Z}_0^{-}} = 
\frac{1-z{\Delta }_0}{1+z{\Delta }_0}
\end{equation}
where ${\Delta }_j:={Z}_j^{-}\left/ {Z}_j^{+}\right. $.

From (\ref{Zhat}), we have
$$
\Delta _{j-1}
  =  \left( \dfrac {\zeta _j \ 
+ z\ \Delta _j}{1 +\zeta _j \ z\ \Delta _j} \right) ^2 
  =  \tau _{j \, , z}({\Delta }_j)
$$ 
and the sequence $\left\{{\Delta }_j\right\}
_{j=0}^M$ satisfies the recurrence relation (\ref{Deltaj}) 
with ${\Delta }_M=1$.  Recall that ${\Delta }_j$ is a random variable
since $\zeta _j= e^{-2\beta \xi _j}$ with $\xi _j$ as given by (\ref{xim}).

\subsection{The Two Point Function}

To compute the numerator
of (\ref{cf}), we repeat the steps in the calculation of the partition
function $Z_M(\xi )$. Our aim is to derive an expression analogous  to
(\ref{ZM}). 

Let $\widetilde{Z}_j=\left( \widetilde{Z}_j^{+},\widetilde{Z}_j^{-}\right) 
$, $j=0,\ldots ,M$, be a sequence of two-component vectors defined
recursively as in the following.

For $j=n_0+1,\,\ldots ,\,M$, with $n_0=$\mbox{dist}$(0,\;x)$, we have 
$\widetilde{Z}_j^\sigma ={Z}_j^\sigma $ , i.e.,
$\widetilde{Z}_j^\sigma$ satisfy the equation (\ref{Zhat}) 
with initial conditions $\widetilde{Z}_M^{+}=\widetilde{Z}_M^{-}=1$.

For $j\leq n_0$, we consider a linear transformation of the form 
\begin{equation}
\label{Ztilde}\widetilde{Z}_{j-1}^\sigma ={Z}_{j-1}^\sigma 
\dfrac{\zeta ^{(1-\sigma )/2}\,\widetilde{Z}_j^{+}+\zeta ^{(1+\sigma )/2}\ z\ 
\widetilde{Z}_j^{-}}{\zeta ^{(1-\sigma )/2}\ {Z}_j^{+}+\zeta
^{(1+\sigma )/2}\,z\ {Z}_j^{-}}\;\; 
\end{equation}
with $\widetilde{Z}_{n_0}^\sigma =\sigma {Z}_{n_0}^\sigma $.

We now observe that the sum over all spin configurations in the numerator of
(\ref{cf}), except by spin at the origin, is determined by
(\ref{Ztilde}) 
and the sum over all spin configurations in the denominator is determined
by (\ref{Zhat}). The two point function (\ref{cf}) can thus be written in
the following form 
\begin{equation}
\label{cf1}\left\langle \sigma _0\sigma _x\right\rangle _M(\xi)=\frac{\ 
\widetilde{Z}_0^{+}-z\ \widetilde{Z}_0^{-}}{{Z}_0^{+}+z\ 
{Z}_0^{-}}. 
\end{equation}

In the following lemma the equation (\ref{cf1}) will be reorganized and
reexpressed in terms of the quantities ${Z}_j^\sigma \; , j = 0, \, 1, \,
\dots , \, M$ and $\sigma = +, \, -$.

\begin{lemma}
{\label{AB}}The sequence of vectors $\widetilde{Z}_n$, $n=1,\ldots ,n_0$,
defined by (\ref{Ztilde}), can be written as 
\begin{equation}
\label{Zt}\widetilde{Z}_{n-1}^\sigma =\left( {\cal A}_n+
\sigma {\cal B}_n\right) {Z}_{n-1}^\sigma ,
\end{equation}
where 
\begin{equation}
\label{Acal}{\cal A}_n=
A_n{\cal B}_{n+1}+\ldots +A_{n_0-1}{\cal B}_{n_0}+A_{n_0}
\end{equation}
and 
\begin{equation}
\label{Bcal}{\cal B}_j=B_jB_{j+1}\ldots B_{n_0} \, ,
\end{equation}
with $j=n,\ldots ,n_0$.
Here $A_j=A_j(\zeta _j ,z)$ and $B_j=B_j(\zeta _j ,z)$, $j=1,\,\ldots ,\,n_0$,
are random variables given by 
\begin{equation}
\label{A}A_j=\frac{\left( z{\Delta }_j\right) ^{-1}-z{\Delta 
}_j}{\left( z{\Delta }_j\right) ^{-1}+z{\Delta }_j+\zeta _j
^{-1}+\zeta _j}\ 
\end{equation}
and 
\begin{equation}
\label{B}B_j=\frac{\zeta ^{-1}-\zeta }{\left( z{\Delta }_j\right)
^{-1}+z{\Delta }_j+\zeta _j ^{-1}+\zeta _j}.
\end{equation}
$\EndofStatement$
\end{lemma}

\noindent {\bf \Proof .} We shall prove Lemma \ref{AB} by
induction. We let $j=n_0$ and observe that, by multiplying the
numerator and the denominator of (\ref{Ztilde}) by $\left(
  \zeta _j^{-(1-\sigma )/2} \left( z{\Delta }_{n_0}\right)^{-1}+
  \zeta _j^{-(1+\sigma )/2}\right) \left/ {Z}_{n_0}^{+}\right. $,
it can be written as

\begin{equation}
\label{Zt1}
\begin{array}{lll}
\dfrac{\widetilde{Z}_{n_0-1}^\sigma  }{{Z}_{n_0-1}^\sigma } & = &
\dfrac{\left( z{\Delta }_{n_0}\right) ^{-1}-z{\Delta }
_{n_0}+\zeta _j^{-\sigma }-\zeta _j^\sigma }{\left( z{\Delta }
_{n_0}\right) ^{-1}+z{\Delta }_{n_0}+\zeta _j ^{-\sigma }+\zeta _j^\sigma } \\
 &   &  \\
 & = & \dfrac{\left( z{\Delta }_{n_0}\right) ^{-1}-z{\Delta }
_{n_0}+\sigma (\zeta _j^{-1} -\zeta _j) }{\left( z{\Delta }
_{n_0}\right) ^{-1}+z{\Delta }_{n_0}+\zeta _j^{-1} +\zeta _j} \\
 &   &   \\
 & = & A_{n_0}+\sigma B_{n_0} .
\end{array} 
\end{equation}

Now, let $j=n+1$. Assuming (\ref{Zt}) valid, (\ref{Ztilde}) can be written
as 
\begin{equation}
\label{Zt2}\frac{\widetilde{Z}_n^\sigma }{{Z}_n^\sigma }=
{\cal A}_{n+1}+\dfrac{\zeta _j^{(1-\sigma )/2}\ -\zeta _j^{(1+\sigma )/2}\
z\  {\Delta }_{n+1}}{\zeta _j^{(1-\sigma )/2}\ +
\zeta _j^{(1+\sigma )/2}\ z\ {\Delta }_{n+1}}{\cal B}_{n+1}. 
\end{equation}
Multiplying the numerator and the denominator of the second term on the right
hand side of this equation by $\zeta _j^{-(1-\sigma
)/2}\left( z{\Delta }_{n+1}\right) ^{-1}+\zeta _j^{-(1+\sigma )/2}$,
gives 
$$
\begin{array}{lll}
\dfrac{\widetilde{Z}_n^\sigma }{\widehat{Z}_n^\sigma }
 & = & {\cal A}_{n+1}+ \dfrac{\left( z{\Delta }_{n+1}\right) ^{-1}-z{\Delta }
_{n+1}+\sigma (\zeta _j^{-1} -\zeta _j) }{\left( z{\Delta }
_{n+1}\right) ^{-1}+z{\Delta }_{n+1}+\zeta _j^{-1} +\zeta _j}{\cal B}_{n+1} \\
 &   &  \\
 & = & {\cal
A}_{n+1}+\left( A_n+\sigma B_n\right) {\cal B}_{n+1} \\
\end{array}
$$
which, in view of (\ref{Acal}) and (\ref{Bcal}), concludes the proof of
Lemma \ref{AB}.
$\Fullbox$

Now we proceed with the proof of Theorem \ref{tpf}. 

\noindent {\bf Proof of Theorem \ref{tpf}.} Using (\ref{Zt}) to
compute (\ref{cf1}), gives 
\begin{equation}
\label{cf2}
\begin{array}{lll}
\left\langle \sigma _0 \sigma _x\right\rangle _M(\xi) & = & 
\dfrac{\left( {\cal A}_1+{\cal B}_1\right) 
{Z}_0^{+}-z\left( {\cal A}_1-{\cal B}_1\right) 
{Z}_0^{-}}{{Z}_0^{+}+z{Z}_0^{-}} \\  &  &  
\\  
& = & {\cal A}_1\dfrac{1-z{\Delta }_0}{1+z{\Delta  }_0}+
{\cal B}_1. 
\end{array}
\end{equation}

The one-point function at $x$ can be computed by a procedure analogous to one
described at the beginning of this subsection. The difference between this
one-point function and the two-point function is the spin at origin. As
before, (\ref{Ztilde}) with 
$j=0, \, \dots , \, n_0$ is of relevance for the description of the numerator 
of $\left\langle \sigma _x\right\rangle$. The iteration gives an expression
of the form (\ref{cf1}) with the minus sign replaced by plus. The one--point
function can thus be written as 
\begin{equation}
\label{opf}\left\langle \sigma _x\right\rangle_M(\xi)=
\frac{\ \widetilde{Z}_0^{+}+z\ \widetilde{Z}_0^{-}}{\
{Z}_0^{+}+z\ {Z}_0^{-}}=
{\cal A}_1+{\cal B}_1\frac{1-z{\Delta }_0}{1+z{\Delta }_0}. 
\end{equation}

Inserting (\ref{cf2}) and (\ref{opf}) into (\ref{t2pfM}) and taking into
account (\ref{sigma1}), yields

\begin{equation}
\label{t2pf1}\left\langle \sigma _0;\sigma _x\right\rangle _M(\xi)=\left[
1-\left( \frac{1-z{\Delta }_0}{1+z{\Delta }_0}\right)
^2\right] {\cal B}_1. 
\end{equation}

When the thermodynamic limit, $M\rightarrow \infty $ has been taken,
the random variables $\Delta _j$'s in (\ref{t2pf1}) can be replaced by
their limits $\displaystyle \lim _{M\rightarrow \infty }\Delta _j$
(recall that ${\cal B}_1$ depends on $\left\{ \Delta _j\right\}
_{j=0}^{n_0}$, each of which defined by a recursion relation with
initial condition $\Delta_M=1$ dependent on the generation $M$). Since
$\Delta _j$ is bounded from above and below, the convergence in
distribution is guaranteed by the convergence of their moments
\cite{Fe}. It follows from Theorem \ref{moments} that
$\displaystyle\lim_{M\rightarrow \infty }\E_\xi \Delta _j$ exists and
is a real analytic function of $z$ in $z\in \R_{+}\setminus \{1\}$.

In order to take the expectation value of (\ref{t2pf1}) we shall use the
following property: for any bounded function 
$f(\xi )=f(\xi _j, \xi^{\prime })$ 
of the random variables $\xi =(\xi _j,\xi ^{\prime })$, we have 
\begin{equation}
\label{EfBj}
 {\E}_\xi f(\xi )B_j(\xi )=p_j{\E}_{\xi^{\prime }}f(1,\xi^{\prime })
 \frac{\zeta^{-1}-\zeta }{\left( z\Delta_j\right)^{-1}+
 z\Delta_j+\zeta ^{-1}+\zeta } 
\end{equation}
since $\zeta _j^{-1}-\zeta _j=e^{2\beta \xi _j}-e^{-2\beta \xi _j}=0$
if 
$\xi _j=0$.

Define 
\begin{equation}
\label{s}
s(x)\; :=\; 1-\left( \frac{1-x}{1+x}\right)^2
\end{equation}
and 
\begin{equation}
\label{t}
t(x)\; =\; t(\zeta ,x)\; :=\; 
\frac{\zeta ^{-1}-\zeta }{x^{-1}+x+\zeta^{-1}+\zeta }.
\end{equation}
Using (\ref{EfBj}) in the expected value of (\ref{t2pf1}) gives
\begin{equation}
\label{expected}
\begin{array}{lll}
{\E}_\xi \left\langle \sigma _0;\ \sigma _x\right\rangle _M(\xi ) & = & 
p_1\ldots p_{n_0}
\; {\E}_{\xi ^{\prime }}s(z\Delta _0)\ \left[
\dprod\limits_{j=1}^{n_0}t(z\Delta _j)\right]  \\  &  &  \\  
& = & \displaystyle \sum_{k=n_0}^{M-1}a_k\ s(w_k)\ \left[
\dprod\limits_{j=1}^{n_0}t(w_{k-j})\right] +p_1\ldots p_M\ s(w_M)\ \left[
\dprod\limits_{j=1}^{n_0}t(w_{M-j})\right] 
\end{array}
\end{equation}
where $w_n=z\tau _z^{(n)}(1),n\in {\N}$. Here, we have first taken
partial expectations with respect to the variables 
$\xi_1,\, \ldots , \, \xi_{n_0}$ and have used in the sequel for all remaining
expectations that the sequence
$\Delta _j$, $j\in \N $, 
satisfies the recurrence relation $z\Delta _{j-1}=zh_j(z\Delta _j)$
with $h_j(1)=1$. Equation (\ref{expect})
then follows since $s(x)$ and $t(x)$ are continuous in $\R_{+}$ and
$w_n$ converges to the solution $w=w(\zeta ,z)$ of the fixed point
equation $w=zh(w)$ in this domain provided $z\in \R_{+}$.
This concludes the proof of Theorem \ref{tpf}.  $\Fullbox $

We now turn to the proof of Theorem \ref{suscept} on the quenched
susceptibility at origin $\chi $.  We note the following facts on the
function $s$ and $t$ (the proof will be omitted):

\begin{proposition}
\label{st}The function $s: \, w\in \R_{+}\longmapsto s(w)\in \R_{+}$ and 
$t: \, (\zeta ,w)\in [0, \, 1]\times \R_{+}\longmapsto t(\zeta ,w)\in \R_{+}$
given by (\ref{s}) and (\ref{t}), respectively, have a maximum value
at $w=1$ with $s(1)=1$ and $t(\zeta ,\, 1)=(1-\zeta )/(1+\zeta )$, are
monotonically increasing function of $w$ in $[0, \, 1]$ and satisfy
$s(w)=s(w^{-1})$ and $t(\zeta ,w)=t(\zeta ,w^{-1})$.
$\EndofStatement$
\end{proposition}

For $z\in [0,\, 1)$ we recall that $w_n,n\in {\N}$, is a monotonically
decreasing sequence with $w_n<1$ and for each $n$, $w_n=w_n(z)$ is
monotonically decreasing function of $z$ in this domain. So, in view
of Proposition \ref{st},
\begin{equation}
\label{ineq1}2t(\zeta ,w_n)\leq 2\frac{1-\zeta }{1+\zeta }<1
\end{equation}
holds for all $\zeta \in (1/3,\, 1]$ and $z\in \R_{+}\left\backslash
  \{1\}\right. $.

In addition, for $\zeta \in [0,\, 1/3)$ (ferromagnetic phase) and $z\in
[0,\, 1)$, $w_n$ converges to a nontrivial solution $w(\zeta ,z)$ of
$w=zh(w)$ (see \cite{BM1}) with $w\nearrow \underline{w}$ as
$z\nearrow 1$ with
$$
\underline{w}=\frac{1-2\zeta -\zeta ^2}{2\zeta ^2}-\frac{1-\zeta ^2}{2\zeta
^2}\left( \frac{1-3\zeta }{1+\zeta }\right) ^{1/2}. 
$$
This and Proposition \ref{st} implies that there exist finite number 
$n_0=n_0(\zeta ,z)$ $\in {\N\,}$ such that for $n>n_0$
\begin{equation}
\label{ineq2}2t(\zeta ,w_n)\leq 2t(\zeta ,\underline{w})=
\frac{2\zeta }{1-\zeta }<1
\end{equation}
holds for all $\zeta \in [0,\, 1/3)$ and $z\in \R_{+}\setminus\{1\}$.

Substituting equation (\ref{expect}) into (\ref{susc}) and taking into
account the inequalities (\ref{ineq1}) and  (\ref{ineq2}), gives
\begin{equation}
\label{susc1}
\begin{array}{lll}
\chi  & = & \dsum\limits_{n=0}^\infty 2^n\dsum\limits_{k=n}^\infty a_k\
s(w_k)\ \left[ \dprod\limits_{j=1}^nt(w_{k-j})\right] +a\
s(w)\dsum\limits_{n=0}^\infty \ \left[ 2t(w)\right] ^n \\  
&  &  \\  
& = & \dsum\limits_{k=0}^\infty a_k\ s(w_k)\dsum\limits_{n=0}^k\ 2^n\left[
\dprod\limits_{j=1}^nt(w_{k-j})\right] +\frac{a\ s(w)}{1-2t(w)}.
\end{array}
\end{equation}

Note that 
$$
\lim \limits_{z\rightarrow 1}\frac{a\ s(w)}{1-2t(w)}\sim C_\eta \left| \eta
\right| \qquad \text{\mbox{as} }\qquad \eta :=\zeta -\zeta_c\rightarrow 0 
$$
holds with $\lim _{\eta \searrow 0}C_\eta =4a/3$ and $\lim _{\eta \nearrow
0}C_\eta =2a/3$. Note also that, for ${\bf p}\in \pi _a$, $0<a\leq 1$, such
that $\dsum_n\,n\, a_n<\infty $, the first term on the second line of 
(\ref{susc1}) does not diverge at $z=1$ for any $\zeta \in [0,\, 1]$ since, by the
dominated convergence theorem,
$$
\lim _{x\nearrow 1}\sum_{k=0}^\infty \frac{1-x^k}{1-x}\,a_k<\infty . 
$$
A non classical exponent $\gamma =\gamma ({\bf p})$
may appear if (\ref{summcond}) is violated.

This concludes the proof of Theorem \ref{suscept}.
$\Fullbox $


\section{The Effective Fugacity and the Analyticity Domain of ${\cal M}_r$}
\label{TheAnalyticityDomainofcalM}
\zerarcounters

In this section we will prove Theorem \ref{moments} concerning the
analyticity domain of ${\cal M}_r$. We split the proof in two
parts, the first, and simplest, dedicated to the function ${\cal
I}_r$  and the second, and more elaborated, to the function ${\cal
F}_r$. The study of the singularities of ${\cal F}_r$ involves a careful
analysis of the dynamical system induced by $ \tau_z$ on the complex
plane. 


Let us first establish the analyticity results on the
functions ${\cal I}_r(z)$, $r\in \N$.

Let us denote by $D_{<1}$ the open unit disk in $S^2$ and by $D_{>1}$ the
complement of its closure. More generally, for $a> 0$ call \be
D_{ < a} := \{w \in S^2 : \; |w| < a \}, \ee
\be
D_{ > a} := \{w \in S^2 : \; |w| > a \}. \ee
For further purposes define also for $a $, $b \in \R_+$, $a < b$, \be
D_{a, \; b} := \{w \in S^2 : \; a < |w| < b \}. \ee

The following  theorem has been proven in \cite{BM1}:

\begin{theorem}
The sequence $\tau _z^{(n)}(1)$, $n\in \N$, of analytic functions on $D_{<1}$
converges uniformly to an analytic function $\tau =\tau (z)$ on the whole
set $D_{<1}$. 
For all $\zeta \in (0,\;1]$ the function $\tau =\tau (z)$ fulfills the fixed
point equation $\tau =h(z\tau )$ on the whole set $D_{<1}$. As a
consequence, $\tau (z)$ has no zeros on $D_{<1}$. 
$\EndofStatement$
\label{analiticidadedasequenciatauz}
\label{remarksobreoszerosdetauz}
\end{theorem}

Analyticity of ${\cal I}_r(z)$ on $D_{<1}$ is, actually, the
statement of Theorem \ref{analiticidadedasequenciatauz}. Let us now
consider ${\cal I}_r(z)$ on $D_{>1}$.  For $r\in \N$, ${\bf p}\in \pi
_a$ and $z\in D_{>1}$, we have
\begin{equation}
{\cal I}_r(z)\ =\ a\ \lim _{n\to \infty }\left[ \tau _z^{(n)}(1)\right] ^r\
=\ a\ \lim _{n\to \infty }\left[ \tau _{z^{-1}}^{(n)}(1)\right] ^{-r}\ =\ a\
\left[ \tau (z^{-1})\right] ^{-r}.
\end{equation}
Therefore, in view of Theorem \ref{remarksobreoszerosdetauz} which says that 
$\tau (w)$ has no zeros for $w\in D_{<1}$, ${\cal I}_r(z)$ is an analytic
function on $D_{>1}$.


\subsection{The Analyticity Domain of ${\cal F}_r$}
\label{TheAnalyticityDomainofFr}

In this subsection we will establish the analyticity results on the
functions ${\cal F}_r(z)$, $r\in \N$. The proof will be given in two 
theorems, according to whether $z\in D_{<1}$ or $z\in D_{>1}$.


Let us prove our first result on the analyticity of ${\cal F}_r(z)$.

\begin{theorem}[The First Analyticity Theorem]
\label{theranaliticityDmenor1}The function ${\cal F}_r(z)$ defined in 
(\ref{calF}) is an analytic function of $z$ in the open set $D_{<1}$ for
all 
${\bf p}\in \pi $ and $\zeta \in [0,\, 1)$.
$\EndofStatement$
\end{theorem}

{\bf \noindent Proof.} The partial sums 
\begin{equation}
\label{calFrN}{\cal F}_{r,\;N}(z):=\sum_{n=0}^N\;a_n\;\left( \tau
_z^{(n)}(1)\right) ^r,\qquad r,\ N\in \N, 
\end{equation}
are analytic in $D_{<1}$ since the pole of $h$ lies in $D_{>1}$. Moreover,
one has $|\tau _z(1)|<1$ for $z\in D_{<1}$ and
so, for any $\varepsilon >0$ one has 
$$
|{\cal F}_{r,\;N}(z)-{\cal F}_{r,\;M}(z)|<\sum_{n\ =\ N+1}^Ma_n\leq
\varepsilon , 
$$
for all $N$ and $M$, $N<M$, large enough, since the sequence 
$\{a_n,\; n\in \N\}$ 
is summable. So, the partial sums form a uniform Cauchy sequence of
analytic functions whose limit exists and is analytic in $
D_{<1}$. $\Fullbox$

We will now prove the analyticity of the function ${\cal F}_r(z)$ in 
$D_{>1}\setminus {\cal Z}$. We start by studying the location of the
singularities of ${\cal F}_r(z)$. First we prove the following lemma.

\begin{lemma}
\label{ywuertfSoqwauera}
Let ${\cal B}\subset \C$ be an open set such that
$\tau _z^{(n)}(1)$, $n\in \N$, are analytic on ${\cal B}$ and 
\begin{equation}
\label{condicaomestra}\inf _{z\in {\cal B}}\;\inf _{n\in \N}\;\left| \tau
_z^{(n)}(1)-\left( \frac{-1}{z\zeta }\right) \right| >0.
\end{equation}
Then ${\cal F}_r(z)$ is an analytic function in ${\cal B}$.
$\EndofStatement$
\end{lemma}

\begin{remark}
Note that $h$ given by (\ref{h}) has a unique double pole at $w=-1/\zeta
$. Moreover, 
$h(w)$ is bounded on $\C \setminus \calO$, where $\calO$ is any open set
containing the pole.
\end{remark}

\begin{remark}
Note that $\inf \limits_{n\in \N}\left| \tau _z^{(n)}(1)-\left(
-z^{-1}/\zeta \right) \right| =0$ is not a necessary condition for ${\cal F}
_r$ to be singular at $z$. For instance, for $z=-1$, one has $\left| \tau
_z^{(n)}(1)-\left( -z^{-1}/\zeta \right) \right| =1/\zeta -1$ for all $n\in 
\N$. However, as we will see below, ${\cal F}_r(z)$ is not analytic at $z=-1$
because $-1$ is an accumulation point of poles of ${\cal F}_r$.
\end{remark}

{\bf \noindent Proof \ of Lemma \ref{ywuertfSoqwauera}.} If condition 
(\ref{condicaomestra}) holds for a given 
${\cal B}$ it implies that there exists
an open neighborhood ${\cal O}$ of $-1/\zeta $ such that, for all $z\in {\cal B}$
and for all $n\in \N$, 
$
z\tau _z^{(n)}(1)\in \C \setminus {\cal O}
$.
Since $\tau _z^{(n+1)}(1)=h(z\tau _z^{(n)}(1))$, this means that there
exists a positive constant $M$, such that for all $z\in {\cal B}$ and all
$ n\in \N$, 
$
|\tau _z^{(n)}(1)|\leq M
$.
Hence, by the same argument used in the proof of Theorem 
\ref{theranaliticityDmenor1}, the partial sums ${\cal F}_{r,\;N}(z)$ form a
uniform Cauchy sequence of analytic functions in ${\cal B}$ and
consequently, their limit exists and is analytic in ${\cal B}$.
$\Fullbox$

Our task now is to localize the points $z\in \C$ where the condition 
\begin{equation}
\label{Kyliuefhgegwffzero}\inf _{n\in \N}\left| \tau _z^{(n)}(1)-\left( 
\frac{-1}{z\zeta }\right) \right| =0 
\end{equation}
holds. The next proposition shows that, for each $n\in \N$, there exist at
least one point $z\in \C$ such that $\left| \tau _z^{(n)}(1)-\left(
-z^{-1}/\zeta \right) \right| =0$.

\begin{proposition}
For each $n\in \N$ there is at least one and at most $2^{n+1}-1$ distinct
solutions in $\C$ of the equation 
\begin{equation}
\label{equacaodassingularidades}\tau _z^{(n)}(1)-\left( \frac{-1}{z\zeta }
\right) =0.
\end{equation}
$\EndofStatement$
\end{proposition}

{\bf \noindent Proof.} We begin by observing that, $\tau
_z^{(n)}(1)$ is the ratio of two polynomials of degree $2^{n+1}-2$: 
$
\tau _z^{(n)}(1)=P_n(z)/Q_n(z) 
$. This is a consequence of the fact that the function $h$ is the
ratio of two polynomials in $z$ of degree $2$. 
To be more explicit, $P_n$ and $Q_n$ are defined recursively through 
\bear
P_{n+1} (z)
& := & \left(\zeta Q_n (z) + z P_n (z) \right)^2 , \qquad n\in\N , 
\nonumber 
\\ Q_{n+1} (z) & := & \left( Q_n (z) + \zeta z P_n (z) \right)^2 , \qquad
n\in\N  
\eear
with $P_0 (z) =  Q_0 (z) :=  1$.

Therefore, (\ref{equacaodassingularidades}) means that 
\begin{equation}
\label{raizesdopolinomio}zP_n(z)+\frac 1\zeta Q_n(z)=0, 
\end{equation}
which has at least one and at most $2^{n+1}-1$ possible distinct solutions
in $\C$, since the left hand side is a polynomial of degree $2^{n+1}-1$.
Clearly, for $n=0$, the unique solution is the pole of $h$, $z=-1/\zeta $.
$\Fullbox$

Let us denote by ${\cal Z}_n$ the finite set of all $z$'s satisfying (\ref
{equacaodassingularidades}) for $n\in \N$. Define 
$$
{\cal Z\;}:=\;\bigcup_{a\in \N}{\cal Z}_a 
$$
and let $\overline{{\cal Z}}$ be the closure of ${\cal Z}$. 
The set difference ${\cal Z}^0:=\overline{{\cal Z}}\setminus 
{\cal Z}$ is the set of accumulation points of ${\cal Z}$.

To establish that ${\cal F}_r$ is holomorphic in 
$\C \setminus \overline{{\cal Z}}$ 
we need to describe where the elements of ${\cal Z}$ and ${\cal Z}^0$ 
are located.

\begin{theorem}
For all $\zeta \in (0,\;1)$ and $k\in \N $ one has ${\cal Z}_k\subset D_{>1}$
with the accumulation points of ${\cal Z}$ lying in the unit circle, i. e., $
\calZ^0\subset S^1$.

In addition, 
$
{\cal Z}_k\cap [1,\;\infty )\;=\;\emptyset  
$,
for all $k\in \N $. The sets ${\cal Z}_k$ are also self-conjugated in
the sense that $\overline{z}\in {\cal Z}_k$ if $z\in {\cal Z}_k$ and, for
all $k\in \N $, one has 
$
{\cal Z}_k\cap (-\infty ,\;-1)\;\neq \;\emptyset 
$.
This last fact together with ${\cal Z}^0\subset S^1$ implies that $-1\in 
{\cal Z}^0$.
$\EndofStatement$
\end{theorem}

{\bf \noindent Proof}. Firstly, observe that the equation 
\begin{equation}
  \label{eqBAsica}
  \tau _z^{(k)}(1)=-\frac 1{z\zeta } 
\end{equation}
has no solutions for $z\in D_1:=D_{<1}\cup S^1$, since $\tau _z^{(k)}(1)$
maps $D_1$ into itself and $\zeta <1$. This equation has no solutions for $
|z|\geq 1/\zeta $ either, since in this region, 
the left hand side of (\ref{eqBAsica}) lies in $D_{>1}$, but the right hand
side in $D_1$. We conclude that 
$
 {\cal Z}\;\subset \;D_{1,\;1/\zeta }
$.
Analogously, (\ref{eqBAsica}) has no solutions in $(1,\;\infty )$, since
there its left hand side is positive while the right hand side is negative.

To prove that ${\cal Z}^0\subset S^1$ we proceed as follows. The
elements of ${\cal Z}_n$ are the poles of $\tau _z^{(n+1)}(1)$ in
$D_{>1}$. Therefore, if $w\in {\cal Z}_n$, $w^{-1}$ is a zero of $\tau
_z^{(n+1)}(1)$ in $D_{<1}$.  The sequence of functions $\tau _z^{(n)}$
converges uniformly in $D_{<1}$ to an analytic function $\tau =\tau
(z)$ (Theorem \ref{analiticidadedasequenciatauz}). According to a
theorem by Hurwitz (see, e.g., \cite{T}) the fact that $\tau $ has no
zeros on $D_{<1}$ (Theorem \ref {remarksobreoszerosdetauz}) implies
that the set of zeros of the functions $ \tau _z^{(n)}$ has no limit
points on $D_{<1}$ and therefore, if they exist, they must lie on
$S^1$.

The sets ${\cal Z}_k$ are all self-conjugated by the definition of
$h$, since the left hand side of (\ref{raizesdopolinomio}) is a
polynomial with real coefficients. This polynomial is also of odd
degree, namely $2^{k+1}-1$.  Both facts together imply that this
polynomial has at least one real root.  However, according to the
previous remarks, the real roots cannot lie in $ [-1,\;\infty )$. So,
they must be negative and lower than $-1$. From the fact that ${\cal
  Z}^0\subset S^1$ it follows that these negative roots must converge
to $-1$ when $k\to \infty $ and so $-1\in {\cal Z}^0$. This shows, in
particular, that ${\cal Z}^0\neq \emptyset $. $\Fullbox$

\begin{remark}
Numerical computations seem to indicate that ${\cal Z}^0\;=\;S^1$ for the
whole ferromagnetic region $0\leq \zeta <1/3$. Unfortunately we were not
able to prove this conjecture.
\end{remark}

After these considerations on the location of the singularities of $\tau
_z^{(n)}$ we turn to the proof of the analyticity statement on ${\cal F}_r$.

\begin{lemma}
\label{lemadauniformidadeemDmaiorqueum}Let ${\cal B}$ be an open set such
that $\overline{{\cal B}}\subset D_{>1}\setminus \overline{{\cal Z}}$, where 
$\overline{{\cal B}}$ is the closure of ${\cal B}$. Then the sequence $\tau
_z^{(n)}(1)$, $n\in \N$, converges to the function $1/\tau (z^{-1})$
uniformly on ${\cal B}$.
\end{lemma}

{\bf \noindent Proof .} For $z\in {\cal B}$ call $w\equiv z^{-1}$ the
``image'' of $z$, and let the open set ${\cal B}_i$ given by
$$
{\cal B}_i\;:=\;\left\{ w\in D_{<1} \mbox{ such that } 
w^{-1}\in {\cal B}\right\}  
$$
be the ``image set'' of ${\cal B}$.

From Theorem \ref{remarksobreoszerosdetauz} we know that $\tau (w)$ has no
zeros on $D_{<1}$. Hence, for $w\in {\cal B}_i$, $|\tau \left( w\right)
|>\rho _1$, for some positive $\rho _1$. Next, we argue that there is also a
positive constant $\rho _2$ such that $|\tau _w^{(n)}(1)|>\rho _2$ for all
$n\in \N$ and all $w\in {\cal B}_i$. For, notice that no element of the
sequence of functions $\tau _w^{(n)}(1)$ has a zero in ${\cal B}_i$. Since
the sequence converges uniformly and the limit function has no zeros
in 
${\cal B}_i$, there must be a positive constant such that $|\tau
_w^{(n)}(1)|>\rho _2$ for all $w\in {\cal B}_i$ and $n\in \N$. The desired
uniformity follows now from 
\begin{equation}
\label{dhbYretnrty}\left| \tau _z^{(n)}-\frac 1{\tau (z^{-1})}\right|
=\left| \frac{\tau (w)-\tau _w^{(n)}(1)}{\tau (w)\;\tau _w^{(n)}(1)}\right|
\leq \frac 1{\rho _1\,\rho _2}\;|\tau (w)-\tau _w^{(n)}(1)|<\varepsilon 
\end{equation}
for any prescribed $\varepsilon >0$, uniformly on ${\cal B}_i$ and for $n$
large enough since, on ${\cal B}_i$, the sequence $\tau _w^{(n)}(1)$
converges uniformly to $\tau (w)$.
$\Fullbox$

\begin{lemma}
\label{lemadaimpossibilidade} Let ${\cal B}$ be as in Lemma \ref
{lemadauniformidadeemDmaiorqueum}. Then the condition 
\begin{equation}
\label{condicaoaserdesmentida}\inf _{z\in \overline{{\cal B}}}\ \inf _{n\in 
\N}\ \left| \tau _z^{(n)}(1)-\left( \frac{-1}{z\zeta }\right) \right| =0
\end{equation}
is impossible for any $\zeta \in (0,\;1]$.
$\EndofStatement$
\end{lemma}

{\bf \noindent Proof.} By the definition of ${\cal B}$, the condition 
$\tau_z^{(n)}(1)+ 1/ (z \zeta) =0$ 
does not hold for any finite $n$. 
Therefore, condition (\ref{condicaoaserdesmentida}) says that 
$$
\inf _{z\in \overline{\calB}\ }\;\left| \frac 1{\tau (z^{-1})}\;-\;\left( 
\frac{-1}{z\zeta }\right) \right| \;=\;0. 
$$
Since $\overline{\calB}$ is closed and $1\left/ \tau (z^{-1})\right.
\;+\;1/( z \zeta ) $ is a continuous function of $z$ on 
$\overline{\calB}$, this implies the existence of a point $u\in 
\overline{\calB}$ such 
that $\tau (u^{-1})=-u\,\zeta $. Since $\tau (w)$ fulfills the
fixed point equation $\tilde \tau =h(w\,\tilde \tau )$ for $w\in D_{<1}$, we
conclude that $-u\,\zeta =h(-\zeta )\equiv 0$. But this is impossible for
$\zeta >0$ and $u\in D_{>1}$.
$\Fullbox$

\begin{theorem}[The Second Analyticity Theorem]
\label{theranaliticityDmenor2}
The function ${\cal F}_r(z)$ defined in (\ref {calF}) is a holomorphic
function of $z$ in the open set $D_{>1}\setminus \overline{\calZ}$ for
all ${\bf p}\in \pi $ and $\zeta \in (0,\,1]$.

If $a_n\neq 0$ the singularities of ${\cal F}_r$ at the points of $\calZ_n$
are poles of order not greater than $\left( 2^{n+1}-1\right) r$.
$\EndofStatement$
\end{theorem}

{\bf \noindent Proof \ of Theorem \ref{theranaliticityDmenor2}.} Pick
an open set $\calB $ such that $\overline{\calB} \subset D_{>1}
\setminus \overline{\calZ}$, where $\overline{\calB}$ is the closure
of $\calB$. Then, analyticity of $\calF_r$ on $\calB$ follows from
Lemma \ref{ywuertfSoqwauera} and Lemma \ref{lemadaimpossibilidade}.
Since $D_{>1}\setminus \overline{\calZ}$ can be covered by such open
sets the theorem is proven.  $\Fullbox$

\subsection{Estimates on the Location of the Poles of $\tau_z^{(n)}(1)$}

To study of how fast the sets $\calZ_k$ accumulate on $S^1$ we will
make use of a contraction theorem, described below, on the inverse
mappings of $h$.

First, some definitions. For $w\in \C$, $w\neq 1/\zeta $, define 
\begin{equation}
\label{gw}
g(w):=\frac{\zeta -w}{\zeta w-1}. 
\end{equation}
and let $u^{1/2}$ be some branch of the square root function in $\C$. Define 
\begin{equation}
\label{hplusminus}h_{+}^{-1}(u)\;:=\;g(u^{1/2}),\qquad
h_{-}^{-1}(u)\;:=\;g(-u^{1/2}). 
\end{equation}
Then one has $(h\circ h_{\pm }^{-1})(z)=z$, $\forall z\in \C$.

\begin{theorem}[Contraction Theorem]
\label{thecontractiontheorem}
There exists a number $\zeta _0$ $\in $ 
$[0,\;1/3)$, whose approximate value is $\zeta _0\simeq 0.29559$, such that
for $\zeta \in [0,\;\zeta _0)$ there exists a strictly positive
function 
$e(\zeta )$ such that for all $u\in D_{1,\;a(\zeta )}$ with 
$
a(\zeta ):=\zeta^{-1} + e(\zeta ) 
$
one has 
\begin{equation}
\label{desigualdademestra1}|h_{\pm }^{-1}(u)|\ <\ |u|.
\end{equation}
$\EndofStatement$
\end{theorem}

\begin{remark}
\label{zetazero}Numerical computations indicate to be impossible to improve
the region of validity of equation (\ref{desigualdademestra1}) to $\zeta
\geq \;\zeta _0$.
\end{remark}

\begin{remark}
As already observed, the inequality (\ref{desigualdademestra1}) becomes an
equality in $S^1$, which is the internal boundary of $D_{1,\;a(\zeta )}$. It
is important to note also that the set $D_{1,\;a(\zeta )}$ contains the 
pole $z=-1/\zeta $, of $h$.
\end{remark}

We will present the proof of the Contraction Theorem in Appendix 
\ref{TheContractionTheorem}. Let us now explore some of its consequences.

\begin{theorem}
\label{urypvneruy}Let $\zeta _0$ as in Theorem \ref{thecontractiontheorem}.
For all $\zeta \in (0,\;\zeta _0)$ and $k\in \N$
\begin{equation}
\label{eruitwyJHHeoiury}
\calZ_k\subset D_{1,\;r_k},
\end{equation}
holds with 
$
  \displaystyle
  r_k=\zeta ^{\frac{-1}{k+1}} 
$.
$\EndofStatement$
\end{theorem}

This theorem illustrates explicitly that the accumulation points
$\calZ^0$ lie in the unit circle and shows how fast the sets $\calZ_k$
converge to it, at least for $\zeta \in (0,\;\zeta _0)$.

{\bf \noindent Proof. }The proof of Theorem \ref{urypvneruy} makes use of
the Contraction Theorem which requires the following technical lemma. We
note that, from equation (\ref{eqBAsica}), there exists a finite sequence of
signals $\left\{ s(l)\in \{-,+\},\;1\leq l\leq k\right\} $ such that 
\begin{equation}
\label{ISDJGHPOIDSUFGS}\frac 1zh_{s(k)}^{-1}\left( \frac
1zh_{s(k-1)}^{-1}\left( \frac 1z\cdots \frac 1zh_{s(1)}^{-1}\left( -\frac
1{z\zeta }\right) \cdots \right) \right) =1. 
\end{equation}

\begin{lemma}
\label{RoihjergtfA}Given $z\in \calZ_k$, $k\in \N$, $k\geq 1$, consider a
sequence of signals $\left\{ s(l)\in \{-,+\},\;1\leq l\leq k\right\} $
satisfying (\ref{ISDJGHPOIDSUFGS}) and define 
$$
w_l:=\frac 1zh_{s(l)}^{-1}\left( \frac 1zh_{s(l-1)}^{-1}\left( \frac
1z\cdots \frac 1zh_{s(1)}^{-1}\left( -\frac 1{z\zeta }\right) \cdots \right)
\right), \quad l\in \{1,\ldots ,k\}. 
$$
Then all $w_l$'s belong to $D_{1,\;a(\zeta )}$, except, of course, $w_k$
which is equal to $1$.
$\EndofStatement$
\end{lemma}

{\bf \noindent Proof.} To prove Lemma \ref{RoihjergtfA} we take, without
loss of generality, $k>1$ and note that, since $z\in D_{1,\,1/\zeta }$, one
has $h_{s(1)}^{-1}\left( -z^{-1}/\zeta \right) \in D_{1,\,1/\zeta }$, by the
Contraction Theorem. Hence, $w_1\in D_{\zeta ,\;1/\zeta }$.

On the other hand $w_1$ cannot belong to $D_{\zeta , \; 1} \cup S^1$ for the
following reason: $h_{\pm}^{-1}$ maps $D_{<1} \cup S^1$ into itself and,
since $1/|z| <1$, this implies $w_2 \in D_{<1}$, which, in turn, would imply
that all subsequent $w_l$'s would also be in $D_{<1}$ and so they could not
reach the point $1$ after a finite number of steps.

Therefore, $w_1\in D_{1,\;1/\zeta }$ and we can apply again the
Contraction Theorem and claim that $h_{s(2)}^{-1}(w_1)\in
D_{1,\;1/\zeta }$. So, we conclude again that $w_2\in D_{\zeta
  ,\;1/\zeta }$. If $k=2$ then, by the hypotheses, $w_2=1$. Otherwise,
if $k>2$, repeating the previous arguments, we conclude again that
$w_2\in D_{1,\;1/\zeta }$. This argument can be repeated a finite
number of times, thus proving the lemma.  $\Fullbox$

Lemma \ref{RoihjergtfA} implies that all $w_l$, $1\leq l<k$, lie in the
region of validity of the Contraction Theorem. Therefore, applying the
inequality (\ref{desigualdademestra1}) repeatedly in (\ref{ISDJGHPOIDSUFGS})
we get 
$
  \zeta |z|^{k+1} < 1
$,
which proves (\ref{eruitwyJHHeoiury}). The fact that $\calZ^0\subset S^1$
follows easily from (\ref{eruitwyJHHeoiury}). This completes the proof of
Theorem \ref{urypvneruy}.
$\Fullbox$

The Contraction Theorem has another consequence.

\begin{theorem}
\label{interseccaovazia}For the sets $\calZ_k$ defined above we have, in the
region of validity of the Contraction Theorem, that 
$
\calZ_l\;\cap \;\calZ_m\;=\;\emptyset 
$,
for all $l$, $m\in \N$, $l\neq m$.
$\EndofStatement$
\end{theorem}

{\bf \noindent Proof.} Let $m>l$ and pick a $z\in \calZ_l\;\cap \;\calZ_m$.
Then one has 
$$
\tau _z^{(m-l)}\left( -\frac 1{z\zeta }\right) \;=\;-\frac 1{z\zeta }. 
$$
This means also that there exists a finite sequence $\left\{ s(a)\in
\{-,+\},\;1\leq a\leq m-l\right\} $ such that 
\begin{equation}
\label{ISDJGHPsydofius2}-\frac 1{z\zeta }=\frac 1zh_{s(m-l)}^{-1}\left(
\frac 1zh_{s(m-l-1)}^{-1}\left( \frac 1z\cdots \frac 1zh_{s(1)}^{-1}\left(
-\frac 1{z\zeta }\right) \cdots \right) \right) . 
\end{equation}
Applying the Contraction Theorem to the last equality we get 
$$
\frac 1{|z|\zeta }<\frac 1{|z|^{m-l+1}\zeta }, 
$$
what means $|z|^{m-l}<1$, a contradiction since $\calZ_l\cap \calZ_m\subset
D_{1,\;1/\zeta }$.
$\Fullbox$



\appendix
\section{The Contraction Theorem}
\label{TheContractionTheorem}
\zerarcounters

This appendix is devoted to the proof of Theorem
\ref{thecontractiontheorem} which is implied by the following theorem.

\begin{theorem}
\label{teoremadog}There exists a $\zeta _0\in [0,\;1/3)$, whose approximate
value is $\zeta _0\simeq 0.29559$, such that there is a strictly positive
function $f(\zeta )$ in the interval $\zeta \in [0,\;\zeta _0)$ such that,
in the open set defined by
$
\displaystyle
1<|w|<\frac 1{\sqrt{\zeta }}+f(\zeta ) 
$, $w \in \C$, 
one has 
$
|g(w)|\;<\;|w|^2
$.
$\EndofStatement$
\end{theorem}

In view of (\ref{gw}), (\ref{hplusminus}) and Theorem
\ref{teoremadog}, for $u\in \C$ such that 
$$
1<|u|<\left( \frac 1{\sqrt{\zeta }}+f(\zeta )\right) ^2=:\frac 1\zeta
+\epsilon (\zeta ), 
$$
one has 
$
|h_{\pm }^{-1}(u)|=|g(\pm u^{1/2})|<|u| 
$.
This concludes the proof of Theorem \ref{thecontractiontheorem}.
$\Fullbox$

{\bf \noindent Proof \ of Theorem \ref{teoremadog}. }Writing $w=x+iy$,
with 
$x$, $y\in \R$, a simple computation shows that  
\begin{equation}
\label{d44POoiusdpfio}|g(w)|^2=\frac{(\zeta ^2+|w|^2)-2\zeta x}{(1+\zeta
^2|w|^2)-2\zeta x}. 
\end{equation}

To analyze the right hand side, consider the following lemma.

\begin{lemma} 
\label{lemadota}
Let $t:\R\longmapsto \R$, be given by $
t(a):= (\alpha -\beta a)(\gamma -\beta a)  $, for $\alpha $, $\beta $
and $\gamma \in \R$ with $\alpha >\gamma >0$ and $ \beta \geq 0$.
Consider the open interval $J:=(-\infty ,\,\gamma /\beta )$. Then, $t$
is a continuous, strictly positive, nowhere constant (for $\beta >0$)
and increasing function in $J$. $\EndofStatement$ 
\end{lemma}

{\bf \noindent Proof.} Continuity on $J$ is obvious. A
computation shows that 
$
t^{\prime }(a)=\beta (\alpha -\gamma )/(\gamma -\beta a)^2 
$.
All claims follow immediately from continuity and from this
relation.  $\Fullbox$

Looking at (\ref{d44POoiusdpfio}) we can make the following identifications: 
\begin{equation}
\label{identificacoes01}\alpha \equiv \zeta ^2+|w|^2;\quad \beta \equiv
2\zeta ;\quad \gamma \equiv 1+\zeta ^2|w|^2;\quad a\equiv x.
\end{equation}
Note that, in view of this 
\begin{equation}
\label{identificacoes02}\alpha -\gamma =(1-\zeta ^2)(|w|^2-1)>0,
\end{equation}
for all $w\in D_{>1}$. Moreover, the condition $a<\gamma /\beta $ means 
$$
x<\frac{1+\zeta ^2|w|^2}{2\zeta }. 
$$
This condition is always satisfied if $|w|<1/\zeta $, since $x\leq |w|$ and
since the stronger condition 
$$
|w|<\frac{1+\zeta ^2|w|^2}{2\zeta } 
$$
is equivalent to the condition $\left( 1-\zeta |w|\right) ^2>0$.

In view of Lemma \ref{lemadota} we conclude that, for $w\in D_{1,\;1/\zeta } 
$, 
$$
|g(w)|^2\leq \frac{\zeta ^2+|w|^2-2\zeta |w|}{1+\zeta ^2|w|^2-2\zeta |w|}=
\frac{(\zeta -|w|)^2}{(1-\zeta |w|)^2}. 
$$
Finally, note that $D_{1,\;1/\sqrt{\zeta }+f(\zeta )}\subset D_{1, \, 1/\zeta }$
for a convenient $f$.

Now we come to the technically most important result of this appendix.

\begin{theorem}
\label{teoremaimportante}Under the assumptions of Theorem \ref{teoremadog}
one has 
\begin{equation}
\label{isuerhgE}\frac{(\zeta -|w|)^2}{(1-\zeta |w|)^2}<|w|^4. 
\end{equation}
$\EndofStatement$
\end{theorem}

{\bf \noindent Proof.} Condition (\ref{isuerhgE}) is equivalent to the
condition that the polynomial 
\begin{equation}
\label{definicaodeP}P(t):=(\zeta -t)^2-t^4(1-\zeta t)^2 
\end{equation}
is strictly negative for $1<t<\frac 1{\sqrt{\zeta }}+f(\zeta )$. Writing it
explicitly one has 
$$
\begin{array}{lll}
P(t) & = & -\zeta ^2t^6+2\zeta t^5-t^4+t^2-2\zeta t+\zeta ^2 \\  
&  &  \\  
& = & \left( (\zeta -t)+t^2(1-\zeta t)\right) \left( (\zeta -t)-t^2(1-\zeta
t)\right) . 
\end{array}
$$
Using the fact that $\pm 1$ is a root of the polynomial $(\zeta -t)\pm
t^2(1-\zeta t)$, we can write 
$$
P(t)=(t-1)(t+1)\left( -\zeta t^2+(1-\zeta )t-\zeta \right) \left( \zeta
t^2-(1+\zeta )t+\zeta \right) . 
$$

This last relation allows us to find explicitly the roots of $P(t)$: they
are the elements of the set $\{-1,\;+1,\;a_{-},\;a_{+},\;b_{-},\;b_{+}\}$,
where 
$$
a_{\pm }\;:=\;\left( \frac{1-\zeta }{2\zeta }\right) \pm \frac 1{2\zeta }
\sqrt{(1+\zeta )(1-3\zeta )} 
$$
and 
$$
b_{\pm }\;:=\;\left( \frac{1+\zeta }{2\zeta }\right) \pm \frac 1{2\zeta }
\sqrt{(1-\zeta )(1+3\zeta )}. 
$$

The following lemma gives the possible signs of the polynomial $P(t)$ on the
real axis.

\begin{lemma}
\label{lemadoszeros}For $a_{\pm }$ and $b_{\pm }$ defined above as functions
of $\zeta $, one has the following results:

\begin{enumerate}
\item  For $0\leq \zeta <1/3$ one has $0\leq a_{-}\leq 1$.

\item  For $0\leq \zeta <1$ one has $0\leq b_{-}\leq 1$.

\item  For $0\leq \zeta <1/3$ one has $b_{+}-a_{+}\geq 1$.

\item  For $0\leq \zeta <\zeta _0$ one has $a_{+}-1/\sqrt{\zeta }>0$,
where $\zeta _0\in [0,\;1/3)$, $\zeta _0\simeq 0.29559$. 
This in particular says that $a_{+}>1$ and, by ${\em 3.}$, $b_{+}>1$.
$\EndofStatement$
\end{enumerate}
\end{lemma}

\begin{figure}[hbtp]
\epsfxsize=13cm
\begin{center}
\leavevmode\epsfbox{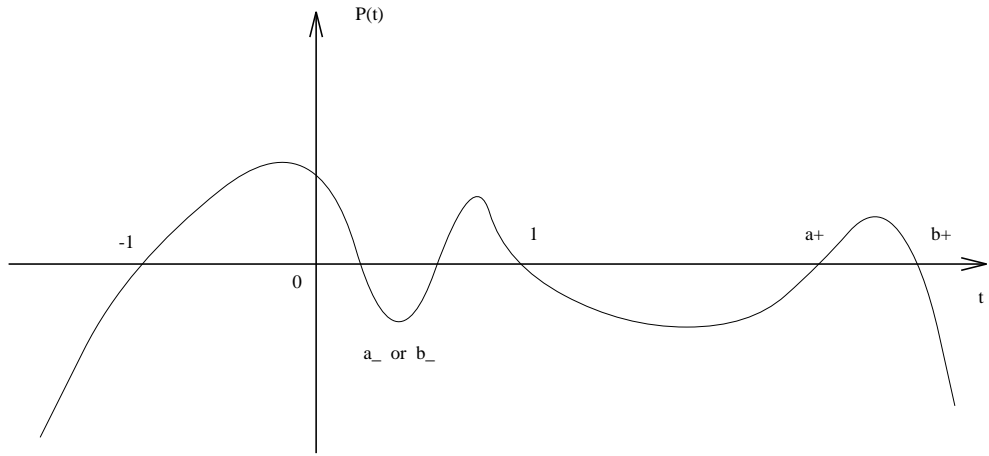}
\end{center}
\caption{The graphic of $P(t)$.}
\label{graficodeP}
\end{figure}

Before we prove this lemma, let us finish the proof of Theorem \ref
{teoremaimportante} and, hence, of the Theorem \ref{teoremadog}. According
to Lemma \ref{lemadoszeros}, $P(t)<0$ if $1<t<a_{+}$. This follows from the
localization of the roots and from the fact that $P$ is a polynomial of even
degree with a negative leading term. See Figure \ref{graficodeP}.

The proof if completed by defining $f(\zeta ):=
a_{+}(\zeta)-1/\sqrt{\zeta }$, which, according with item $4$ of Lemma
\ref{lemadoszeros}, is strictly positive for $0\leq \zeta <\zeta _0$.
This completes also the proof of the Contraction Theorem.  $\Fullbox$


{\bf \noindent Proof of Lemma \ref{lemadoszeros}.} We will prove each of the
items of Lemma \ref{lemadoszeros} separately.

{\em \noindent Proof \ of item 1.}

The hypothesis that $a_{-}\geq 0$ is equivalent to 
$
(1-\zeta )^2\geq (1+\zeta )(1-3\zeta ) 
$
for $0\leq \zeta <1/3$. This last relation means $4\zeta ^2\geq 0$, which is
obviously verified.

The hypothesis that $a_{-}\leq 1$ is equivalent to 
$$
1-\sqrt{(1+\zeta )(1-3\zeta )}\leq 3\zeta  ,
$$
which is equivalent to 
$
(1-3\zeta )^2\leq 1-2\zeta -3\zeta ^2
$.
This means $4\zeta (3\zeta -1)\leq 0$, what is true for $0\leq \zeta <1/3$,
the equality holding only if $\zeta =0$.

{\em \noindent Proof \ of item 2.}

The hypothesis that $b_{-}\geq 0$ is equivalent to  
$
(1+\zeta )^2\geq (1-\zeta )(1+3\zeta )
$,
which is equivalent to $4\zeta ^2\geq 0$, which is, of course, always true.

The hypothesis that $b_{-}\leq 1$ is equivalent to 
$
(1-\zeta )^2\leq (1-\zeta )(1+3\zeta )
$.
This means that $4\zeta (\zeta -1)\leq 0$, which is always true for $0\leq
\zeta \leq 1$.

{\em \noindent Proof \ of item 3.}

Since 
$$
b_{+}-a_{-}=1+\frac 1{2\zeta }\left( \sqrt{1+2\zeta -3\zeta ^2}-
\sqrt{1-2\zeta -3\zeta ^2}\right) , 
$$
item ${\em 3}$ is proven, provided the term between parenthesis
above is positive. This is implied by the inequality $1+2\zeta
-3\zeta^2 \geq 1-2\zeta -3\zeta ^2$, which is always true for $\zeta
\geq 0$.

{\em \noindent Proof \ of item 4.}

The condition $a_{+}>1\left/ \sqrt{\zeta }\right. $ means that 
\begin{equation}
\label{tiuertyertrtrt}\sqrt{(1+\zeta )(1-3\zeta )}>\zeta +2\sqrt{\zeta }-1. 
\end{equation}
The right hand side is strictly negative for 
$0\leq \zeta <(\sqrt{2}-1)^2\simeq 0.171$. 
So, in this region the condition above is automatically
satisfied. On the other hand, if the right hand side of
(\ref{tiuertyertrtrt}) 
is positive, we can square both sides and arrive to the equivalent
condition 
\begin{equation}
\label{poldeg3}4s(s^3+s^2+s-1)<0 
\end{equation}
where $s=\sqrt{\zeta }$. The polynomial $s^3+s^2+s-1$ has one real
root at $s_0\simeq 0.543689$ and two complex roots at $s_{\pm }\simeq
-0.77\pm 1.115i$.  We call $\zeta _0:=s_0^2$, which gives $\zeta
_0\simeq 0.295597$. Thus, condition (\ref{poldeg3}) is satisfied for
$0<\zeta <\zeta _0$. With this the proof of Lemma \ref{lemadoszeros}
is complete.  $\Fullbox$



\end{document}
